\documentclass[journal]{IEEEtran}
\usepackage{latexsym,epsf,times,amsmath,color,amssymb,indentfirst,fancyhdr,colortbl,
bm}
\usepackage{subcaption}
 \usepackage{graphicx}
\newcommand{\nc}{\newcommand}
\nc{\be}{\begin{equation}}
\nc{\ee}{\end{equation}}
\def\calA{{\mathcal{A}}}

\def\calG{{\mathcal{G}}}

\def\calI{{\mathcal{I}}}
\def\calK{{\mathcal{K}}}
\def\calL{{\mathcal{L}}}
\def\calN{{\mathcal{N}}}

\def\calV{{\mathcal{V}}}

\def\setC{{\mathbb{C}}}

\def\setR{{\mathbb{R}}}

\nc{\Abf}{\mathbf{A}}
\nc{\Cbf}{\mathbf{C}}
\nc{\Dbf}{\mathbf{D}}
\nc{\Fbf}{\mathbf{F}}
\nc{\Gbf}{\mathbf{G}}
\nc{\Hbf}{\mathbf{H}}
\nc{\Ibf}{\mathbf{I}}
\nc{\Lbf}{\mathbf{L}}
\nc{\Nbf}{\mathbf{N}}
\nc{\Qbf}{\mathbf{Q}}
\nc{\Rbf}{\mathbf{R}}
\nc{\Sbf}{\mathbf{S}}
\nc{\Ubf}{\mathbf{U}}
\nc{\Wbf}{\mathbf{W}}
\nc{\Xbf}{\mathbf{X}}
\nc{\Ybf}{\mathbf{Y}}
\nc{\Zbf}{\mathbf{Z}}
\nc{\abf}{\mathbf{a}}
\nc{\bbf}{\mathbf{b}}
\nc{\hbf}{\mathbf{h}}
\nc{\gbf}{\mathbf{g}}
\nc{\nbf}{\mathbf{n}}
\nc{\pbf}{\mathbf{p}}
\nc{\qbf}{\mathbf{q}}
\nc{\rbf}{\mathbf{r}}
\nc{\sbf}{\mathbf{s}}
\nc{\ubf}{\mathbf{u}}
\nc{\vbf}{\mathbf{v}}
\nc{\xbf}{\mathbf{x}}
\nc{\ybf}{\mathbf{y}}
\nc{\zbf}{\mathbf{z}}
\nc{\unobf}{\mathbf{1}}
\nc{\zerobf}{\mathbf{0}}
\nc{\vbfs}{\mathbf{\scriptsize v}}
\nc{\Phibf}{\mathbf{\Phi}}
\nc{\Psibf}{\mathbf{\Psi}}
\nc{\Thetabf}{\mathbf{\Theta}}
\nc{\Lambdabf}{\mathbf{\Lambda}}
\nc{\Sigmabf}{\mathbf{\Sigma}}
\nc{\Omegabf}{\mathbf{\Omega}}
\nc{\xibf}{{\mbox{\boldmath $\xi$}}}
\nc{\rhobf}{{\mbox{\boldmath $\rho$}}}
\nc{\xibfs}{{\mbox{\boldmath \scriptsize $\xi$}}}
\nc{\diag}{\text{diag}}
\nc{\sign}{\text{sign}}
\nc{\E}{\text{E}}
\def\defeq{\stackrel {\scriptscriptstyle \Delta}{=}}

\IEEEoverridecommandlockouts
\makeatletter
\title{Graph Based Imaging for Synthetic Aperture Radar}
\author{\IEEEauthorblockN{Shahzad Gishkori and Bernard Mulgrew}
\thanks{S. Gishkori and B. Mulgrew are with Institute for Digital Communications (IDCOM), The School of Engineering, The University of Edinburgh, UK. 
Emails: \{s.gishkori, bernie.mulgrew\}@ed.ac.uk}
\thanks{This work was supported by Jaguar Land Rover and the UK-EPSRC grant EP/N012240/1 as part of the jointly funded Towards Autonomy: Smart and Connected Control (TASCC) Programme.}
}
\providecommand{\keywords}[1]{\textbf{\textit{Index terms---}} #1}
\begin{document}
\maketitle
\vspace{-0.3in}
\begin{abstract}
\noindent
In this paper, we propose graph signal processing based imaging for synthetic aperture radar.
We present a modified version of fused least absolute shrinkage and selection operator to cater for graph structure of the radar image.
We solve the cost function via alternating direction method of multipliers.
Our method provides improved denoising and resolution enhancing capabilities.
It can also accommodate the compressed sensing framework quite easily.
Experimental results corroborate the validity of our proposed methodology.
\end{abstract}
{\keywords Graph Signal Processing, SAR imaging, Fused Lasso, ADMM}
\section{Introduction}
\label{sect:intro}
Synthetic aperture radar (SAR) \cite{Carrara_95,Cumming_SAR_2005} is known to provide all-weather high-resolution images. This has lead to its rampant use in a variety of applications including surveillance, automation and  medical imaging.
Generally, SAR operates in two modes to provide high cross-range resolution, i.e., stripmap mode (Strip-SAR), where a target scene is illuminated at a fixed aspect angle and the radar traverses over the aperture, and spotlight mode (Spot-SAR), where a target scene is illuminated from different aspect angles over the aperture \cite{Jakowatz_SpotSAR_1996,Soumekh_SAR_SP_99}. Our focus in this paper is on Spot-SAR.  However, our proposed techniques are applicable for both the modes. A large body of work is available to enhance the quality of SAR images in terms of denoising and super-resolution. Most of the proposed techniques have been borrowed from imaging in optical sensors. 
Nonetheless, enhancing the quality of a SAR image is a challenging task. One of the reasons is the disparity between range- and the cross-range resolution, with latter being smaller than the former. This leads to an image spread over an irregular grid. Secondly, radar returns from a target scene are heavily dependent upon the aspect angles and/or position of radar. Small variations in the aspect angles or position can produce completely different reflectivity pattern which results in a nonuniform image. This can be challenging in imaging extended objects where adjacent reflective points on the object may produce drastically different reflectivities. 
Thus, a straightforward application of general imaging techniques on SAR provides limited gains. 
However, one of the qualities of SAR, that differentiates it from other imaging sensors, is the availability of precise ranging information. 
Exploiting this extra information can potentially enhance the quality of a SAR image, as shown in this paper.
\\
Graph signal processing (GSP) \cite{Shuman_GSP_2013,DSPG_2013} has recently been proposed as a technique which processes signals lying on specific data structures defined by the graphs. This essentially means that all elements/samples of the signal form vertices on a graph and the edge weights connecting these vertices provide a measure of similarity between them \cite{Sandryhaila_GraphSampl_2015,Geert_GraphSampl_2016}. Thus, a graph signal can assume any irregularity of structure and it can get processed accordingly. 
In our case, different range- and cross-range resolutions give rise to an irregular grid structure of a SAR image, which is further complicated by overlapping grids from different aspect angles.
Therefore, substantial gains can be obtained by applying the GSP techniques for SAR imaging.
\\
Fused least absolute shrinkage and selection operator (FLasso) \cite{fusedLasso,G_icassp14} is known to provide element-wise sparsity as well as smoothness. We have recently used FLasso in \cite{G_Radar18} for SAR imaging of an automotive scene for improved azimuth resolution. In FLasso, smoothness is achieved by total variation (TV) \cite{Rudin_NTV}.
TV is an edge-preserving norm and it has been at the forefront of image denoising for many years. The basic idea is to minimise the difference between consecutive image pixels which results in noise reduction and feature enhancement. TV can be related to a graph with unit edge weights between adjacent pixels only. Recently, some works have advocated the use of nonlocal neighbours for improved results \cite{Buades_2005,Osher_NLTV_2008,NLTV_Lou_2010}, i.e., a nonlocal TV (NLTV). Nonlocal neighbourhood is defined in terms of similarity of patches centred around different pixels over the complete image. The reference image is generally a coarse estimate of the reconstructed image. The edge weights are then a function of a Euclidean distance between the patches. NLTV provides good results. However, searching for neighbours is a computationally intensive process. In \cite{Mahmood_GSP_2018}, NLTV is used in the context of GSP for tomographic reconstruction, where the search over neighbours is reduced by using $\calK$-nearest neighbours algorithm and the edge weights are updated adaptively. 
However, in NLTV, apart from the computational complexity issues of searching for nonlocal neighbours, edge weights are still dependent upon pixel intensities. Given the nonuniform reflectivity pattern of SAR images, generating edge weights based on pixel intensities can provide limited gains only. 
In this paper, we propose a new definition of neighbourhood for SAR images. We call it extended neighbourhood (EN). It essentially comprises of all the pixels within a certain proximity to the reference pixel. The neighbourhood is defined in terms of ranges between the scatterers. Thus, the weight function reflects the actual ranges. In this way, the requirement of an exhaustive search for neighbours is removed and the nonuniform nature of the reflectivity pattern is also taken care of (especially for the extended targets). 
The reason is that for an extended object, there is a high probability of finding similar scatterers in close proximity. 
Also, given the precise range information in SAR images, such an approach can be quite effective. 
In the light of above, we combine the concept of GSP with EN and reformulate the FLasso cost function, named as graph fused Lasso with extended neighbourhood total variation (GFL-ENTV). We compare our approach with a number of methods, including the NLTV approach. For a fair comparison, we cast NLTV in GFL framework, i.e., GFL-NLTV.
Our method can easily accommodate the compressed sensing (CS) \cite{donoho, candes} framework as well. 
This is particularly useful in the case of insufficient SAR measurements. Therefore, we provide a composite signal model, accordingly. 
\\
\noindent
{\bf \em Contributions}.
We propose a graph based formulation of FLasso. We propose the concept of extended neighbourhood which is defined in terms of actual SAR ranges of the target scene. The weight function obtained in this respect has reduced computational complexity and is better suited to tackle the problems of irregularity of spatial grid and nonuniformity of reflectivity in SAR images.
We solve our cost function via alternating direction method of multipliers (ADMM) \cite{Boyd_admm_2011,BerTsit}, which enjoys the benefits of parallelisation and fast convergence. 
Our proposed approach results in enhanced spatial resolution and improved SAR imaging. We provide experimental results to prove that our proposed method outperforms a number of other imaging techniques.
\\
\noindent
{\bf \em Notations}.
Matrices are in upper case bold while column vectors are in lower case bold,
$(\cdot)^T$ denotes transpose whereas $(\cdot)^H$ denotes Hermitian,
$[\abf]_{i}$ is the $i$th element of $\abf$
and 
$[\Abf]_{ij}$ is the $ij$th element of $\Abf$,
$\hat{\abf}$ is the estimate of $\abf$,
$\defeq$ defines an entity,
$|\calA|$ denotes the cardinality of set $\calA$,
and 
the $\ell_p$-norm is denoted as $||\abf||_p = (\sum_{i=1}^{N} |[\abf]_{i}|
^p)^{1/p}$.
\begin{figure}[tb]
\centering
\includegraphics[scale=0.6]{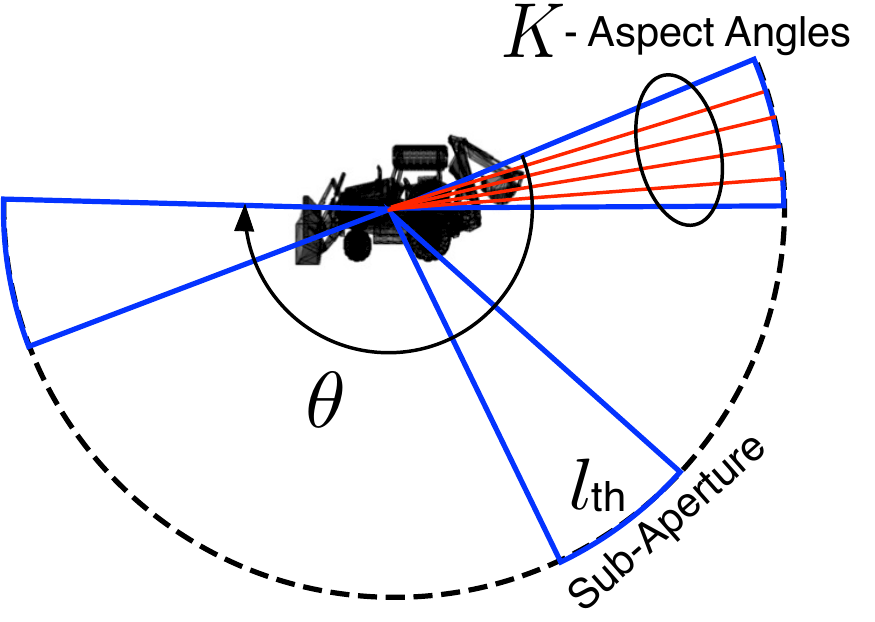} 
\caption{Spot-SAR Measurement Schematic}
\label{fig:backhoe_schem}
\end{figure}
\section{Signal Model}
\label{sect:sig_mod}
In Spot-SAR, the target scene is illuminated from different aspect angles $\theta$, which form the synthetic aperture. Depending on the range of aspect angles, synthetic aperture can be narrow or wide. 
In contrast to the wide-angle, a narrow-angle synthetic aperture assumes that the target reflectivity is isotropic over all aspect angles. However, a wide-angle synthetic aperture can be modelled to consist of many narrow-angle synthetic apertures, named as sub-apertures. Figure \ref{fig:backhoe_schem} shows the measurement schematic of such a wide-angle Spot-SAR.
In Spot-SAR, the received signal (after some post-processing) can be modelled as a spatial Fourier transform of the target field reflectivity (see \cite{Cetin_SAR_2014} and references therein), i.e., 
\be
r(\gamma_m,\theta_k^l) = \sum_{n=1}^{N} s(x_n,y_n;\theta_k^l) \exp\left(-j2\pi \gamma_m \phi_{n,k}^l \right) + n(\gamma_m,\theta_k^l) 
\label{eq:rec_mod}
\ee
where $\phi_{n,k}^l \defeq 2(x_n\cos\theta_k^l + y_n\sin\theta_k^l)/c$, $\gamma_m$ is the $m$th  spatial frequency, for $m=1,\cdots,M$, $\theta_k^l$ is the $k$th aspect angle, for $k=1,\cdots,K$, within $l$th sub-aperture, for $l=1,\cdots,L$, $s(x_n,y_n;\theta_k^l)$ is the reflectivity function of the $n$th spatial location $(x_n,y_n)$ in a Cartesian coordinate system, conditioned on $\theta_k^l$, for $n=1,\cdots,N$ and $n(\gamma_m,\theta_k^l)$ is the additive Gaussian noise corresponding to $\gamma_m$ and $\theta_k^l$.
Now, we can write (\ref{eq:rec_mod}) in the following discrete form.
\be
\rbf_k^l = \Phibf_k^l \sbf_k^l + \nbf_k^l
\label{eq:rec_mod_disc}
\ee
where $\rbf_k^l$ is an $M\times 1$ vector of samples of $r(\gamma_m,\theta_k^l)$, $\Phibf_k^l$ is an $M\times N$ matrix of the samples of $\exp(-j2\pi \gamma_m \phi_{n,k}^l)$, $\sbf_k^l$ is an $N\times 1$ vector of samples of field reflectivity function $s(x_n,y_n;\theta_k^l)$ and $\nbf_k^l$ is an $M\times 1$ vector of samples of  noise $n(\gamma_m,\theta_k^l)$.
Note, all the aforementioned samples are taken for a given $\theta_k^l$. 
Now, a composite model of (\ref{eq:rec_mod_disc}) can be written as
\be
\ybf^l = \Thetabf^l \sbf^l + \vbf^l
\label{eq:rec_mod_comp}
\ee
where $\ybf^l \defeq \Psibf [\rbf_1^{l\,T},\cdots,\rbf_{K}^{l\,T}]^T$ is a $KJ\times 1$ vector, $\Thetabf^l \defeq \Psibf  [\Phibf_1^{l\,T},\cdots,\Phibf_{K}^{l\,T}]^T$ is a $KJ\times N$ matrix, $\vbf^l \defeq \Psibf [\nbf_1^{l\,T},\cdots,\nbf_{K}^{l\,T}]^T$ is a $KJ\times 1$ vector and $\Psibf$ is a $KJ\times KM$ random selection matrix, with $J\le M$.
Note, the above model is valid for narrow-angle sub-apertures, i.e., $K$ is spread over few degrees of angles, under the assumption that the reflectivity function $s(x_n,y_n;\theta_k^l)$ remains isotropic over all $k$ for a given $l$. Thus, $\sbf^l = \sbf_k^l, \, \forall k$.
After finding an estimate of $\sbf^l$, $\forall l$, in (\ref{eq:rec_mod_comp}), a composite response to the field reflectivity of the $n$th spatial location $(x_n,y_n)$ can be obtained by the following simple metric.
\be
[\tilde{\sbf}]_n = \max_l |[\hat{\sbf}^l]_n|^2
\label{eq:s_n}
\ee
for $n=1,\cdots,N$. We can see that (\ref{eq:s_n}) essentially finds a peak reflectivity response of the $n$th spatial location among all sub-apertures.
Note, we solve (\ref{eq:rec_mod_comp}) for each $l$th sub-aperture, independently, and drop the superscript depicting sub-aperture in subsequent sections, for notational simplicity. 
\section{GSP Based SAR Imaging}
\label{sec:gsp}
A graph can be defined as a tuple $\calG\defeq (\calV,w)$, where $\calV \defeq \{v_1, \cdots, v_{N} \}$ is a set of $N$ vertices and $w$ is a weight map between each pair of elements in $\calV$, i.e., $w(v_n,v_{n'})\in \setR_+$, where $v_n,v_{n'} \in \calV$. Generally, $w(v_n,v_{n})=0$, i.e., no self-loops. Note, in this paper, we consider undirected graphs, i.e., $w(v_n,v_{n'})=w(v_{n'},v_n)$. Two vertices are connected to each other if their respective weight map is nonzero. For an $n$th vertex, all its connected vertices define its neighbourhood $\calN_n$, i.e., $\calN_n\defeq \{v_{n'}\in\calV : w(v_n,v_{n'}) \neq 0 \}$. The weight map $w$ can be described in the form of an $N\times N$ adjacency matrix $\Wbf$, where $[\Wbf]_{nn'} = w(n,n')$. An $N\times N$ degree matrix $\Dbf$ is defined as, $[\Dbf]_{nn} = \sum_{n'} w(v_n,v_{n'})$, which is a diagonal matrix. Then, the (combinatorial) graph Laplacian is defined as $\Lbf = \Dbf - \Wbf$.
\\
As explained in Section \ref{sect:intro}, radar signals can be processed under the GSP framework. Thus, a radar graph signal $\sbf$ can be defined as a map from graph vertices to complex-valued signal samples, i.e., 
$
\sbf: \;\; \calV \to \setC, \;\; v_n \mapsto [\sbf]_n
$.
Transforming a graph signal by the graph Laplacian generates weighted smoothing of the graph signal, i.e., 
\be
[\Lbf \sbf]_n = \sum_{[\sbf]_{n'} \in \calN_n} [\Wbf]_{nn'} \left( [\sbf]_n - [\sbf]_{n'} \right)
\label{eq:Ls}
\ee
which shows that the GSP framework enables processing variations of a signal spread over any kind of graph structure, as determined by $\calN_n$.
Now, in the context of GSP, our proposed GFL optimisation problem can be written as
\be
\hat{\sbf} = \mathop{\arg \min}\limits_{\small \sbf} \frac{1}{2} \|\ybf-\Thetabf\sbf\|_2^2 + \lambda_e\|\sbf\|_1^1  
+ \lambda_f \|\Lambdabf\sbf\|_1^1
\label{eq:gfl}
\ee
where $\lambda_e,\lambda_f > 0$ are penalty parameters for element-wise sparsity and graph 
fusion\footnote{
Note, in the case of complex valued signals, some authors, e.g., \cite{Cetin_2001}, suggest fusing/smoothing only the magnitude part out, instead of both real and imaginary parts, since the phase is assumed to be random \cite{Munson_randPhase_1984}. However, in our view, the random phase is a constraint of the measurement system and not necessarily a requirement of fusing complex values. Therefore, in the present paper, we fuse both the real and imaginary parts. Future extensions of the work may include the random phase constraints as well.
}, respectively,
and $\Lambdabf$ is the $\sum_{n=1}^{N}|\calN_n| \times N$ graph difference matrix defined as $\Lambdabf \defeq [\Lambdabf_1^T, \cdots, \Lambdabf_{N}^T]^T$, where $\Lambdabf_n$ is an $|\calN_n|\times N$ matrix such that
\be
 [\Lambdabf_n]_{ij}=
   \begin{cases}
 	+[\Wbf]_{n \{\calN_n\}_ i}	&j=n \\
	-[\Wbf]_{n \{\calN_n\}_ i}	&j = \{\calN_n\}_ i\\
    0	&\text{otherwise}
   \end{cases}
	\label{eq:Lam_casz}
\ee
where (with some abuse of notation) $\{\calN_n\}_ i$ denotes the vertex index of the $i$th element in set $\calN_n$, for $i=1,\cdots, |\calN_n|$, and $j = 1,\cdots,N$. 
From (\ref{eq:Lam_casz}), we can see that $\Lambdabf_n$ is in fact a reshaped form of the nonzero elements of the $n$th row of $\Lbf$, i.e., $[\Lbf]_{n:}\longrightarrow \Lambdabf_n$.
Thus, the fusion part of the GFL can be expanded as
\be
\left\| \Lambdabf \sbf \right\|^1_1 = \sum_{n=1}^{N}  \sum_{[\sbf]_{n'}  \in \calN_n} [\Wbf]_{nn'} \left\|  \left( [\sbf]_n - [\sbf]_{n'} \right) \right\|_1^1
\label{eq:Lambda_s}
\ee	
which creates parsimony over the weighted absolute difference of the neighbouring spatial samples. Thus, GFL encourages sparsity both in the individual elements of $\sbf$ as well as in neighbouring pairs of the elements of $\sbf$. This problem formulation results in increased resolution of the target scene as well as improved imaging of the extended targets.
We solve the GFL problem via ADMM. Thus, (\ref{eq:gfl}) can be re-written as
\begin{align}
[\hat{\sbf},\hat{\ubf},\hat{\zbf}] = \mathop{\arg \min}\limits_{\small \sbf, \ubf, \zbf} 
&\frac{1}{2} \|\ybf-\Thetabf\sbf\|_2^2 + \lambda_e\|\ubf\|_1^1 + \lambda_f \|\zbf\|_1^1 \notag \\
&\text{s.t.} \;\;\; \ubf=\sbf, \;\; \zbf = \Lambdabf\sbf
\label{eq:GLF_0}
\end{align}
where $\ubf$ and $\zbf$ are $N\times 1$ and $\sum_{n=1}^{N}|\calN_n| \times 1$ auxiliary variables, respectively. Now, the cost function in (\ref{eq:GLF_0}) can be written in the following unconstrained form.
\begin{align}
&\calL(\sbf,\ubf,\zbf,\rhobf_u,\rhobf_z) = \frac{1}{2} \|\ybf-\Thetabf\sbf\|_2^2 + \lambda_e\|\ubf\|_1^1 + \lambda_f \|\zbf\|_1^1 +\notag \\
&\rhobf_{u}^H (\ubf-\sbf) + \frac{c_u}{2}\|\ubf-\sbf\|_2^2 + \rhobf_z^H(\zbf-\Lambdabf\sbf) + \frac{c_z}{2}\|\zbf-\Lambdabf\sbf\|_2^2
\label{eq:GFL_L}
\end{align}
where $\rhobf_u$ and $\rhobf_z$ are Lagrange multipliers, and $c_u$ and $c_z$ are positive constants.
An iterative solution of (\ref{eq:GLF_0}), for the $t$th iteration can be obtained by minimising (\ref{eq:GFL_L}) over $\sbf$, $\ubf$ and $\zbf$, one-at-a-time, while keeping other variables fixed.
Thus, a closed-from estimate of $\sbf$ can be written as
\begin{align}
&\hat{\sbf}^{[t]} = \left(\Thetabf^H\Thetabf + c_u \Ibf + c_z\Lambdabf^T\Lambdabf \right)^{-1} \notag \\
&\times \left( \Thetabf^H\ybf + \hat{\rhobf}_u^{[t-1]}  + c_u\hat{\ubf}^{[t-1]}  + \Lambdabf^T\hat{\rhobf}_z^{[t-1]}  + c_z\Lambdabf^T\hat{\zbf}^{[t-1]}  \right).
\label{eq:s_est}
\end{align}
Note, the matrix inversion in (\ref{eq:s_est}) does not depend on iteration index $t$. Therefore, its off-line calculation can save substantial amount of computation. Also, matrix inversion lemma can be used to further reduce the computation. An estimate of $\ubf$ can be written as
\be
\hat{\ubf}^{[t]} = \eta \left( \left[ \hat{\sbf}^{[t-1]} - \dfrac{\hat{\rhobf}_u^{[t-1]}}{c_u} \right], \dfrac{\lambda_e}{c_u} \right)
\label{eq:u_est}
\ee
where $\eta(\sbf,\lambda) = \sign(\sbf)(|\sbf|-\lambda)_+$, with $\sign([\sbf]_n)=[\sbf]_n/|[\sbf]_n|$, and an estimate of $\zbf$ can be written as
\be
\hat{\zbf}^{[t]} = \eta \left( \left[\Lambdabf \hat{\sbf}^{[t-1]} - \dfrac{\hat{\rhobf}_z^{[t-1]}}{c_z} \right], \dfrac{\lambda_f}{c_z} \right).
\label{eq:z_est}
\ee
The Lagrange multipliers can be updated as
\begin{align}
\hat{\rhobf}_u^{[t]} &= \hat{\rhobf}_u^{[t-1]} + c_u(\hat{\ubf}^{[t]}-\hat{\sbf}^{[t]})
\label{eq:rho_u} \\
\hat{\rhobf}_z^{[t]} &= \hat{\rhobf}_z^{[t-1]} + c_z(\hat{\zbf}^{[t]}-\Lambdabf\hat{\sbf}^{[t]}).
\label{eq:rho_z}
\end{align}
Now, the weights in the adjacency matrix are generally obtained from a Gaussian kernel, i.e.,
\be
 [\Wbf]_{nn'}=
   \begin{cases}
    \exp\left( - \dfrac{\Delta_{nn'}^2}{2\sigma^2} \right)	&\text{if} \;\; \Delta_{nn'}\le D \\
    0	&\text{otherwise}
   \end{cases}
    \label{eq:w_gaus}
\ee
where $\sigma^2$ is the variance and $\Delta_{nn'}$ is a function of physical or feature space distances between vertices $[\sbf]_n$ and $[\sbf]_{n'}$. In NLTV, $\Delta_{nn'}$ is the Euclidean distance between image patches of certain dimension, centred around the neighbouring vertices. 
Generally, a coarse estimate of the reconstructed image is used to find these weights. Thus, $\Delta_{nn'}$ is defined as
\be
\Delta_{nn'} \defeq \left\| [\hat{\sbf}]_{\calI_n} - [\hat{\sbf}]_{\calI_{n'}} \right\|_2
\label{eq:delta_nltv}
\ee
where ${\calI_n}$ is a set of indices corresponding to the pixels in image patch centred around vertex $[\sbf]_n$.
In case of radar, actual ranges of different scatterers on the target scene are available. Therefore, we propose to use these ranges in defining EN. Thus, $\Delta_{nn'}$ can be defined as
\be
\Delta_{nn'} \defeq 
\left\| 
\begin{pmatrix}
x_n \\
x_{n'}
\end{pmatrix}
-
\begin{pmatrix}
y_n \\
y_{n'}
\end{pmatrix}
\right\|_2
\label{eq:delta_gfl}
\ee
where $(x_n,y_n)$ and $(x_{n'},y_{n'})$ correspond to spatial locations of vertices $[\sbf]_n$ and $[\sbf]_{n'}$, respectively. Comparing (\ref{eq:delta_nltv}) and (\ref{eq:delta_gfl}), we can see that the adjacency matrix $\Wbf$ needs to be updated for each sub-aperture due to the former, whereas, $\Wbf$ is calculated only once, due to the latter. Thus, our proposed method (using (\ref{eq:delta_gfl})) can guarantee substantial reduction in the computational complexity.
\section{Experimental Results}
\label{sec:sim}
\begin{figure}[t]
\centering
\begin{subfigure}[b]{0.49\linewidth}
    \centering
     \includegraphics[width=.99\textwidth]{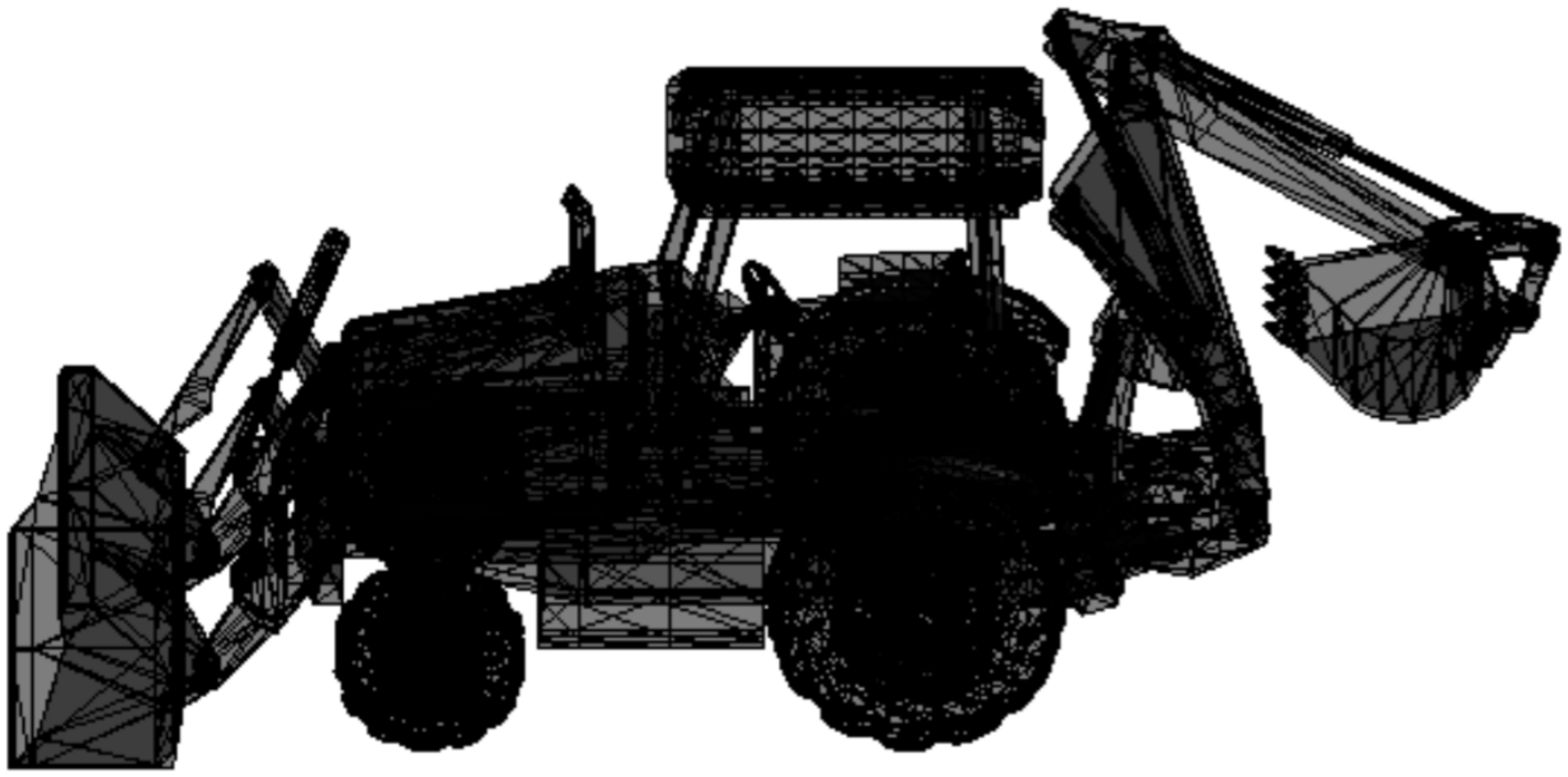} 
    \caption{}\label{fig:backhoe}
  \end{subfigure}
  \begin{subfigure}[b]{0.49\linewidth}
    \centering
     \includegraphics[width=.99\textwidth]{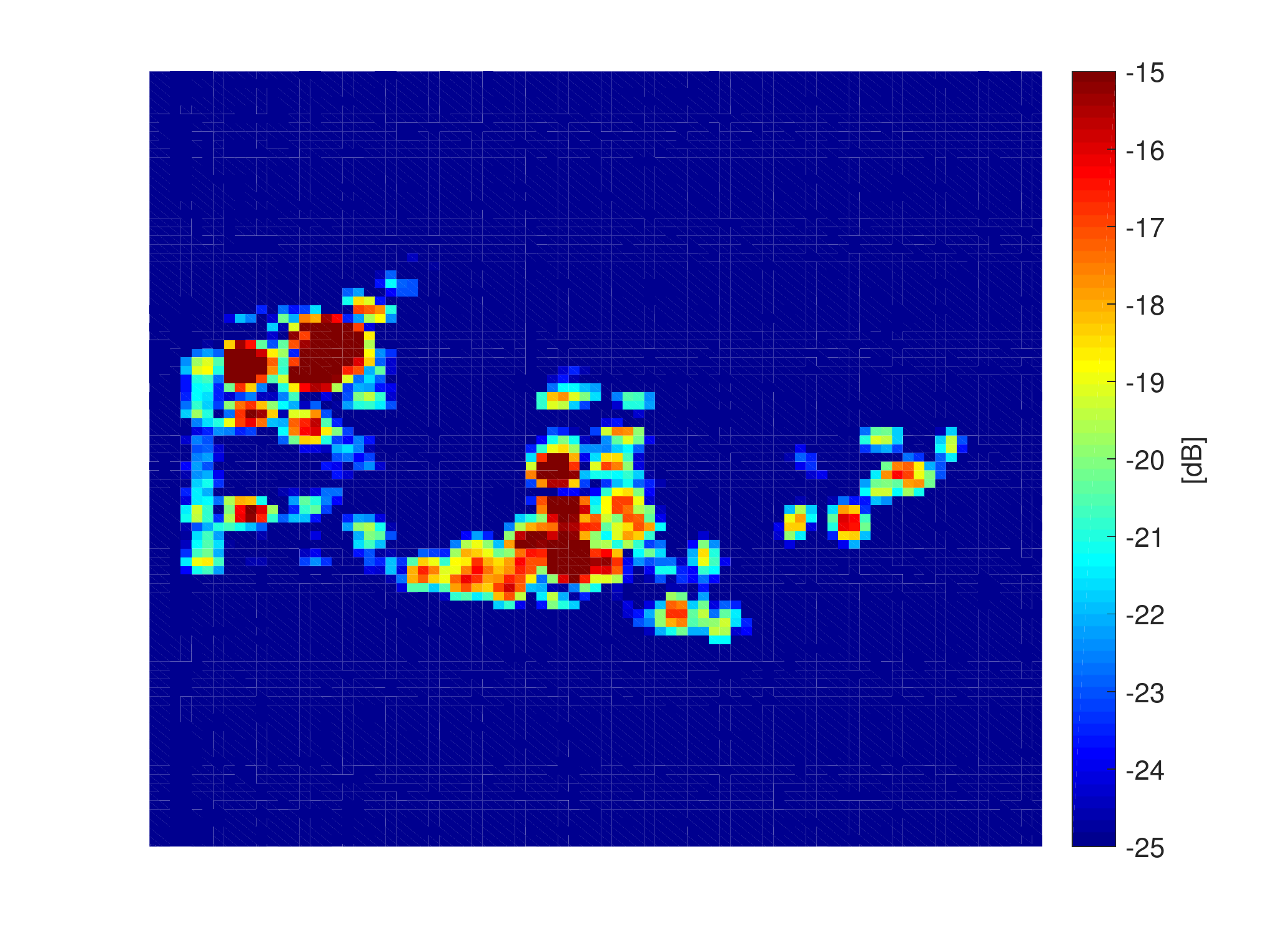} 
    \caption{}\label{fig:mf}
  \end{subfigure}
   \begin{subfigure}[b]{0.49\linewidth}
    \centering
     \includegraphics[width=.99\textwidth]{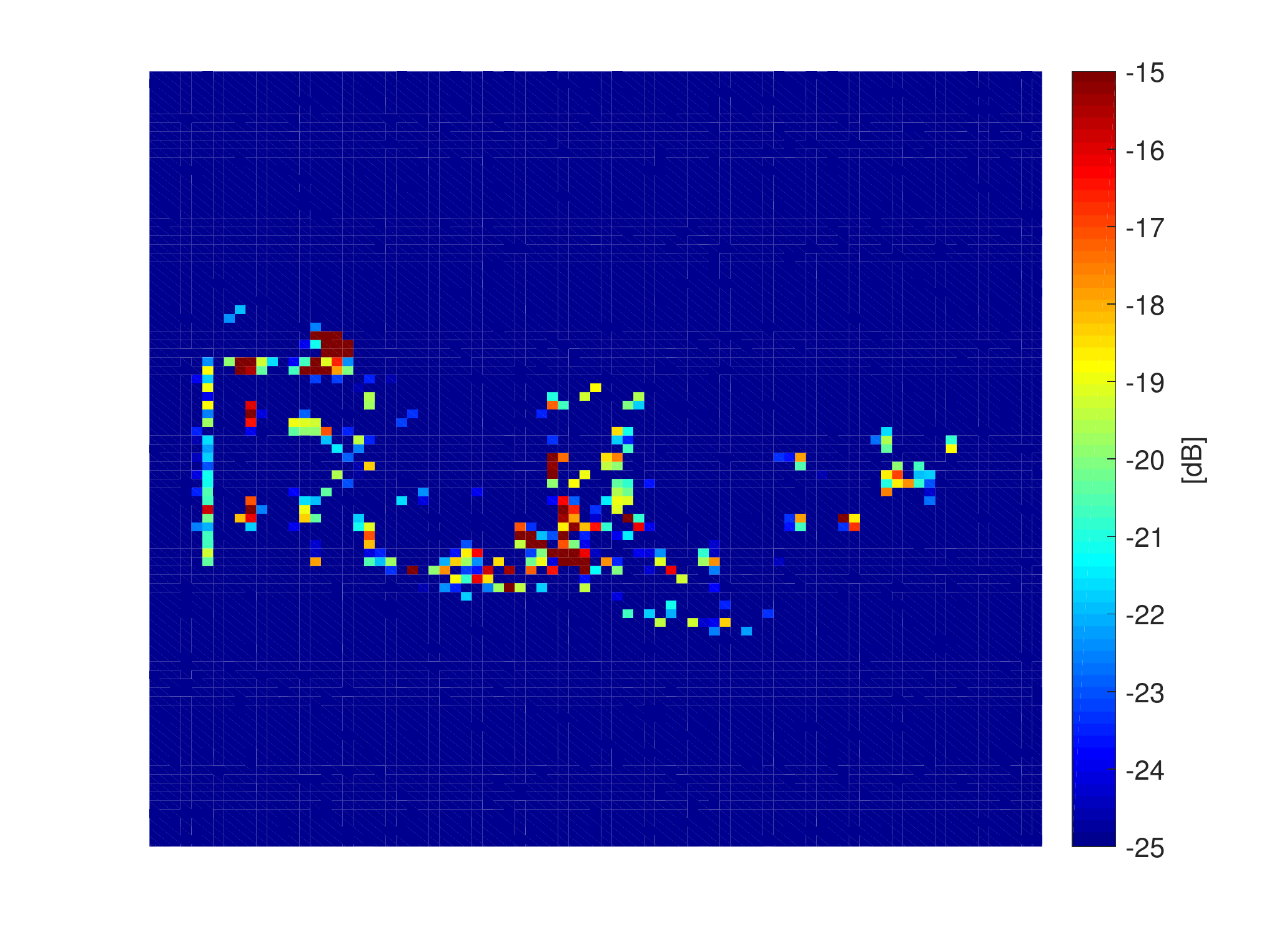} 
    \caption{}\label{fig:nltv}
  \end{subfigure} 
   \begin{subfigure}[b]{0.49\linewidth}
    \centering
     \includegraphics[width=.99\textwidth]{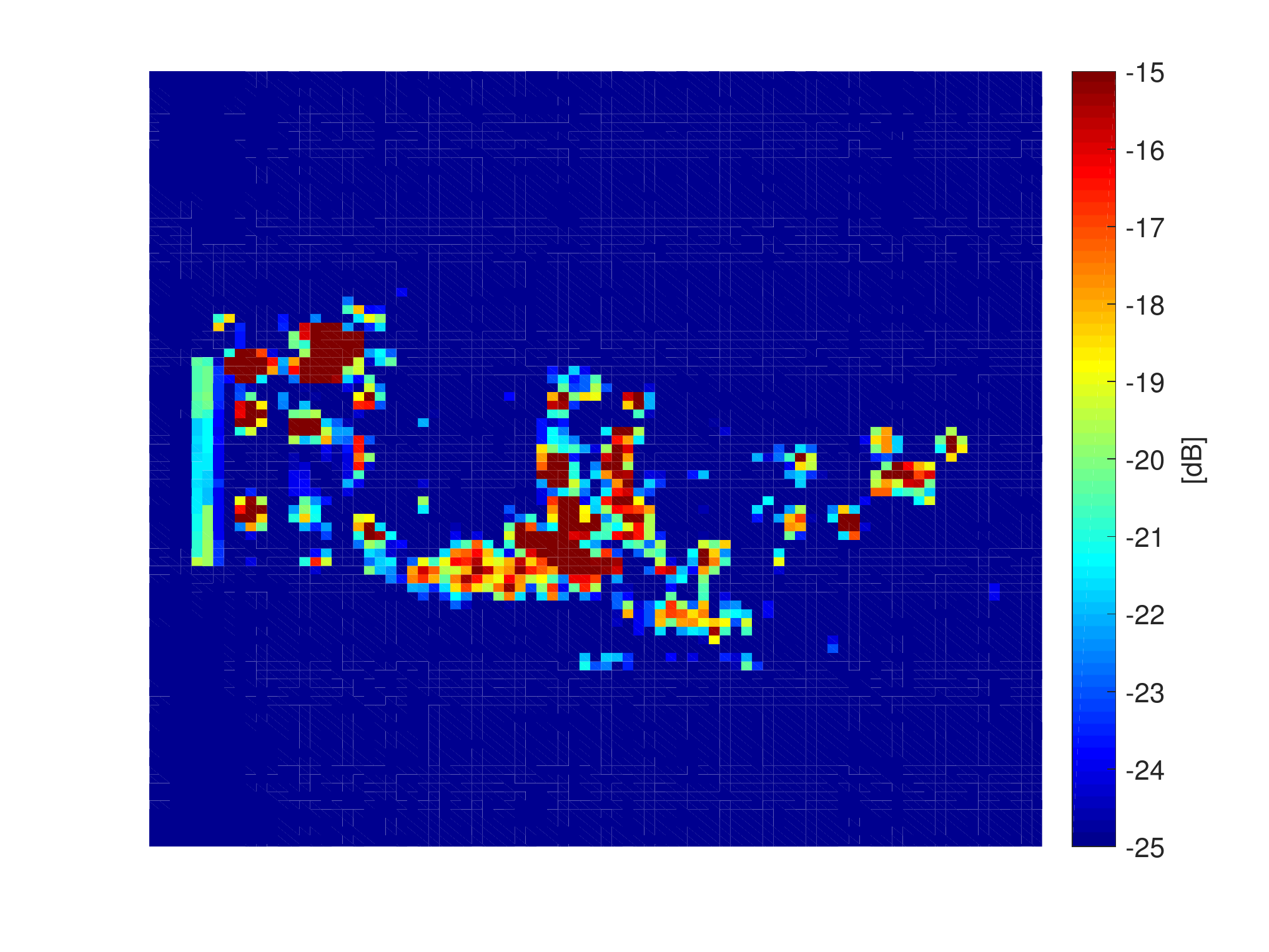} 
    \caption{}\label{fig:tv}
  \end{subfigure}
  \begin{subfigure}[b]{0.49\linewidth}
    \centering
     \includegraphics[width=.99\textwidth]{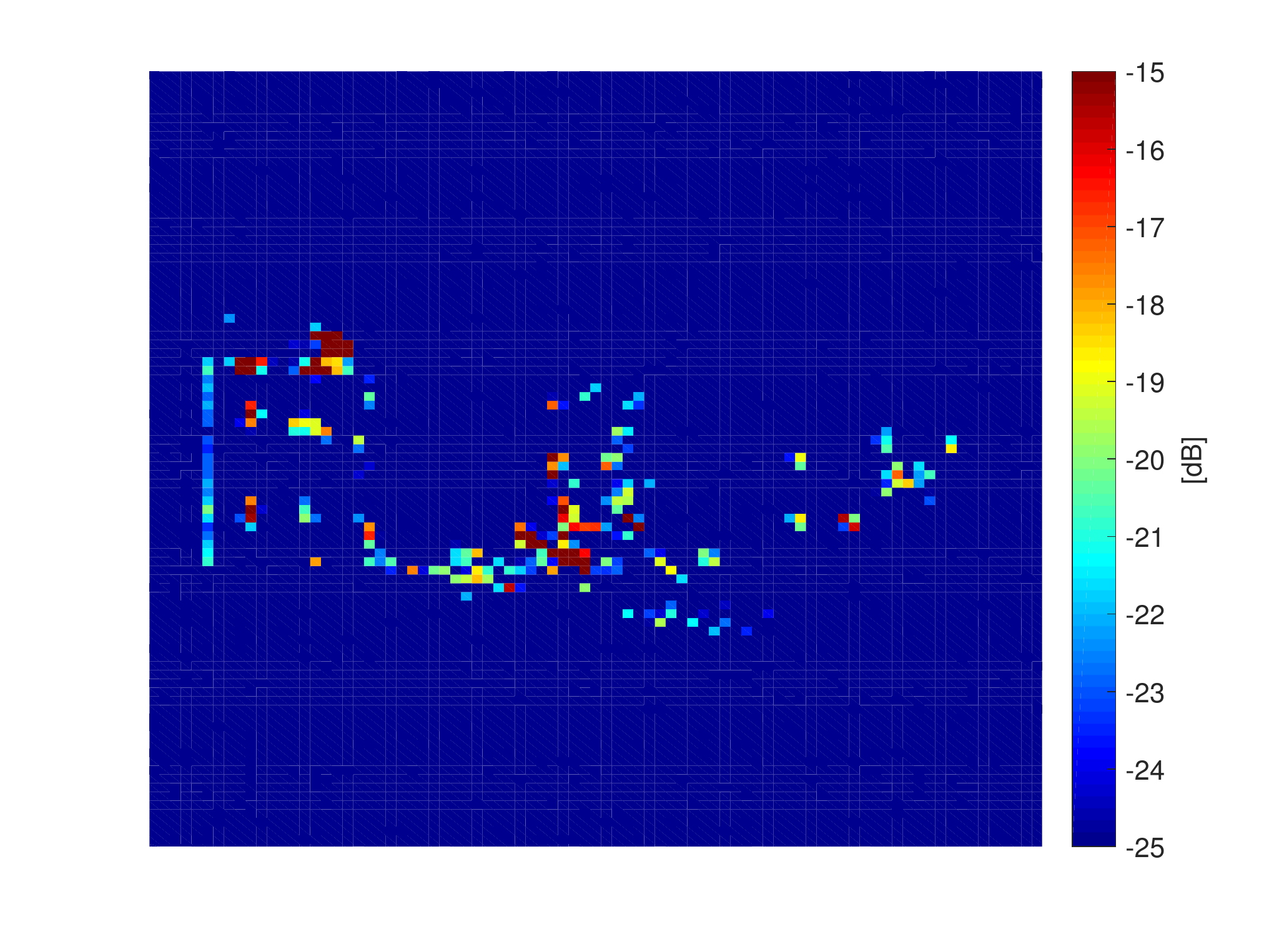} 
    \caption{}\label{fig:fltv}
  \end{subfigure}
  \begin{subfigure}[b]{0.49\linewidth}
    \centering
     \includegraphics[width=.99\textwidth]{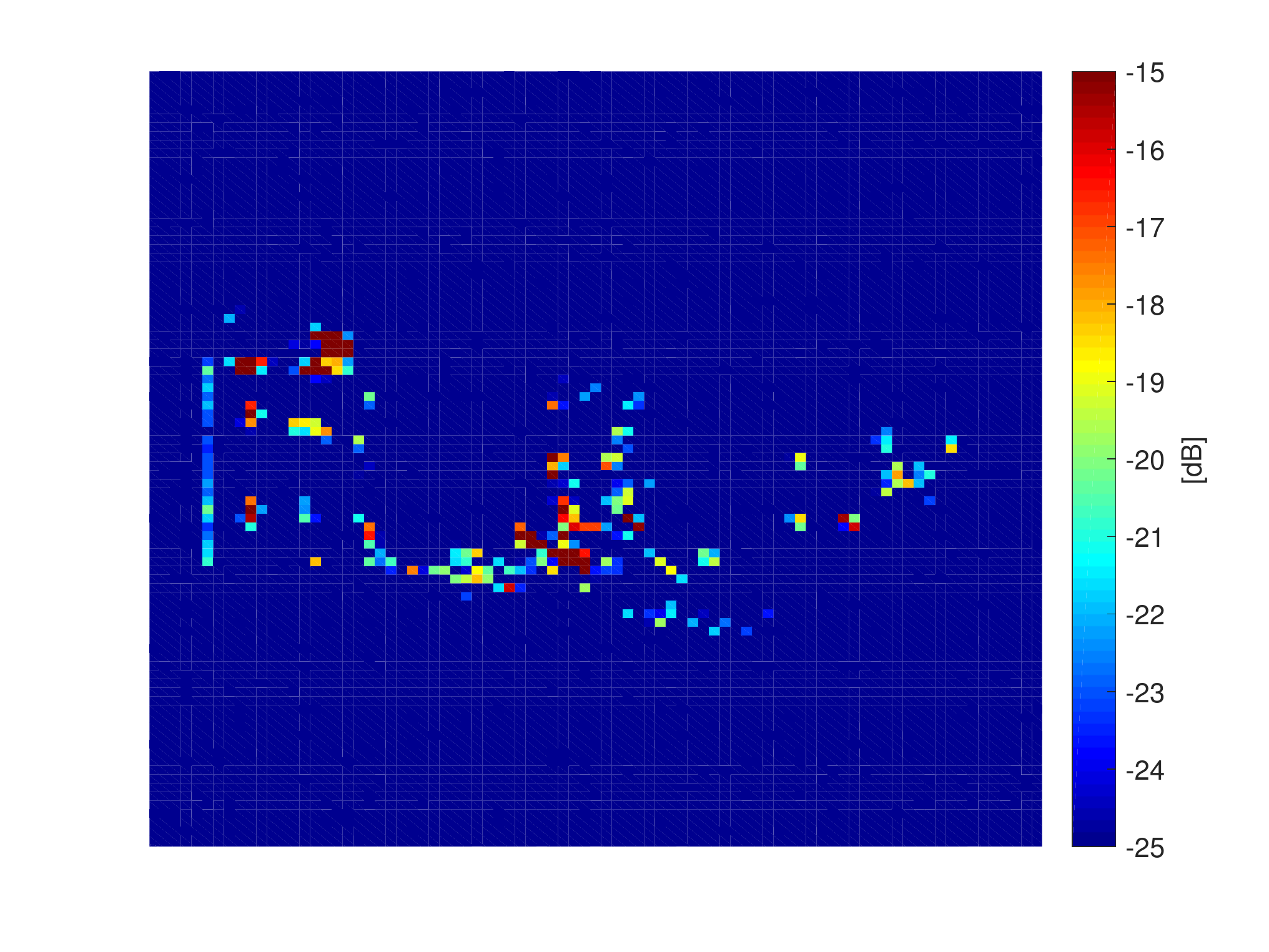} 
    \caption{}\label{fig:fltv_pt75}
  \end{subfigure}
  \begin{subfigure}[b]{0.49\linewidth}
    \centering
     \includegraphics[width=.99\textwidth]{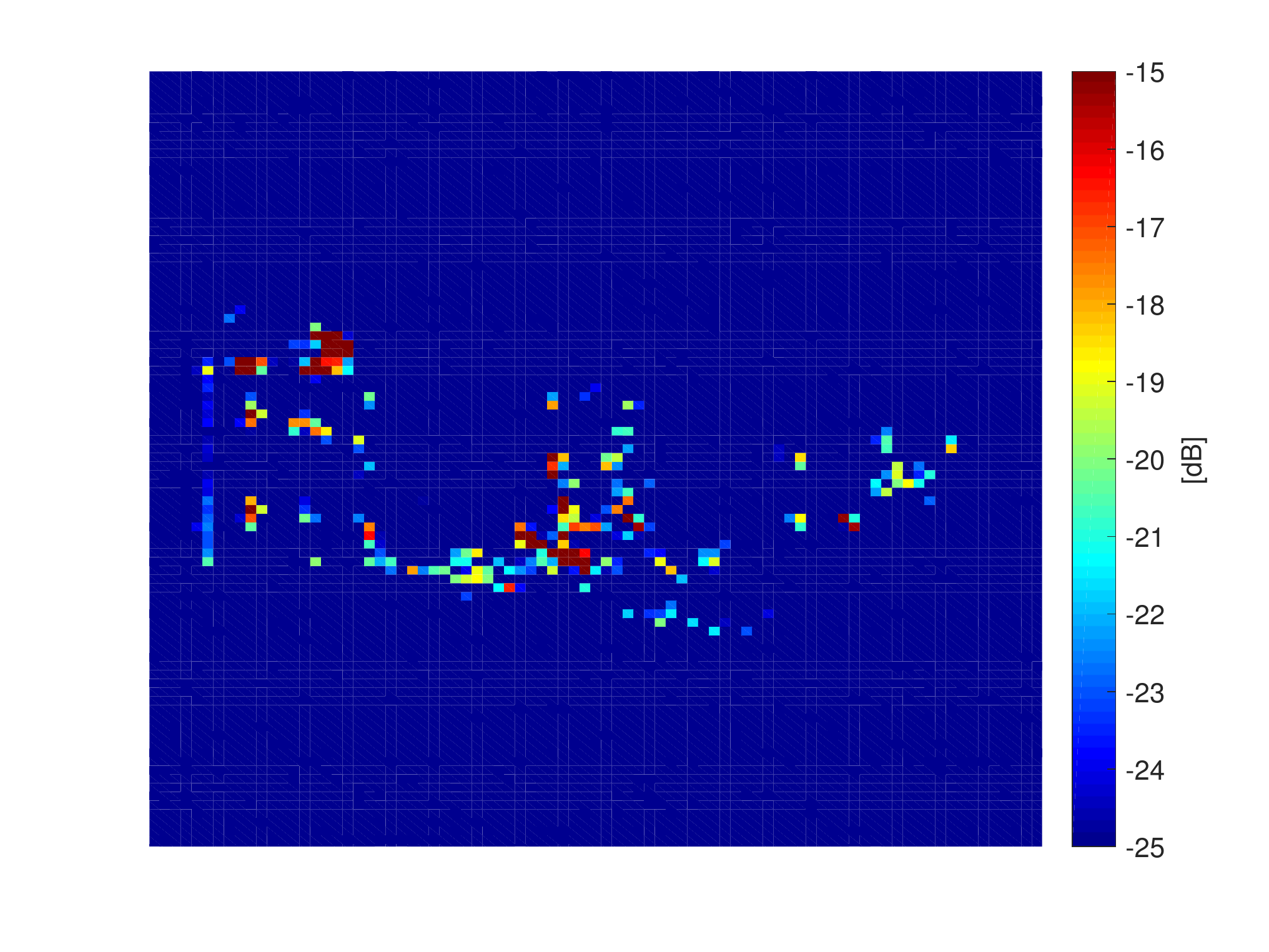} 
    \caption{}\label{fig:fltv_pt5}
  \end{subfigure}
  \begin{subfigure}[b]{0.49\linewidth}
    \centering
     \includegraphics[width=.99\textwidth]{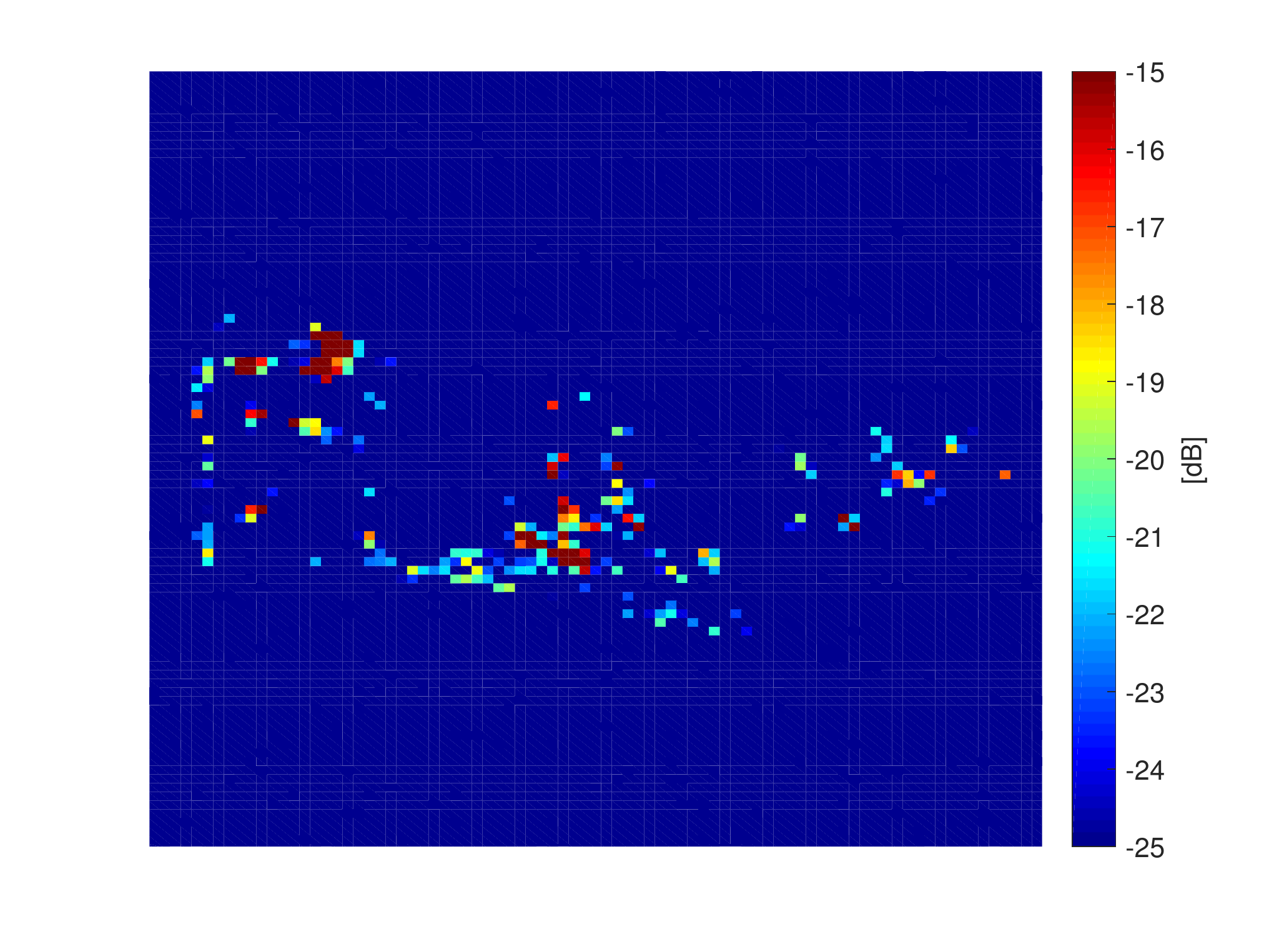} 
    \caption{}\label{fig:fltv_pt25}
  \end{subfigure}
  \caption{
  Performance Comparison of SAR Imaging Techniques:
  (\subref{fig:backhoe}) Backhoe target,
  (\subref{fig:mf}) BP,
  (\subref{fig:nltv}) GFL-NLTV,
  (\subref{fig:tv}) $2$D-TV,
  (\subref{fig:fltv}) GFL-ENTV,
  (\subref{fig:fltv_pt75}) GFL-ENTV with $75\%$ frequency samples,
    (\subref{fig:fltv_pt5}) GFL-ENTV with $50\%$ frequency samples,  
      (\subref{fig:fltv_pt25}) GFL-ENTV with $25\%$ frequency samples
  }
  \label{fig:targets}
\end{figure}

For experiments, we consider the dataset of a backhoe target \cite{backhoe_mstar}. The dataset has been synthetically generated as a dome over the target at an elevation angle of $30^\circ$, for the angular range $\theta \in [-10^\circ,100^\circ]$, with a bandwidth of $5.9$ GHz centred at a frequency of $10$ GHz. 
Figure \ref{fig:backhoe} shows the target.
We divide the complete angular range into $L=22$ sub-apertures, where each $l$th sub-aperture covers an angular range of $5^\circ$, comprising of $K=70$ angular samples. Instead of using the complete frequency bandwidth, we restrict ourselves to a bandwidth of $0.5$ GHz, which generates $M=44$ frequency samples.
We reconstruct the target scene as a grid of $128\times 128$ cells/pixels, which generates $N=16384$ spatial image samples.
We compare the performance of a number of methods for SAR imaging. In this respect, we reconstruct the target scene for each sub-aperture and then use (\ref{eq:s_n}) to construct the final image. The most common method of SAR imaging is back projection (BP). Since our signal model (\ref{eq:rec_mod}) maps the spatial locations directly into the measurements, a BP solution essentially reduces to a matched filtering solution. Figure \ref{fig:mf} shows the performance results of BP based imaging. We can see that the bright scatterers are smeared with each other, causing a reduced spatial resolution.
Figure \ref{fig:tv} shows the performance results of $2$D-TV. We can see some improvement in resolution. However, the bright scatterers do not show a large contrast.
Figure \ref{fig:nltv} shows the performance results of GFL-NLTV. For a fair comparison we have used the GFL framework, i.e., (\ref{eq:s_est})-(\ref{eq:rho_z}), where the stopping criterion is an update tolerance of $10^{-5}$ or the maximum iterations of 100. However, weights of the adjacency matrix have been obtained via (\ref{eq:delta_nltv}) in (\ref{eq:w_gaus}). Parameter $D$ in (\ref{eq:w_gaus}) has been selected so that the neighbourhood search window for each pixel is $21\times 21$ and set $\calI_n$ in (\ref{eq:delta_nltv}) has been designed to represent indices of a $3\times 3$ image patch centred around the $n$th pixel. For each sub-aperture, we use a BP based image as an estimate of $\sbf$ in (\ref{eq:delta_nltv}). The performance results of GFL-NLTV show improved resolution capabilities where the bright scatterers are clearly visible. However, few spurious pixels can also be seen in the reconstructed image. 
A major negative aspect of this technique is the increased computational complexity. 
Figure \ref{fig:fltv} shows the performance results of GFL-ENTV based imaging. Weights of the adjacency matrix have been obtained from (\ref{eq:delta_gfl}) in (\ref{eq:w_gaus}). These weights are the same for all of the sub-apertures. Then, the SAR image is obtained by iterating over (\ref{eq:s_est})-(\ref{eq:rho_z}).
Note, the stopping criterion and spatial parameters of the Gaussian kernel are the same as GFL-NLTV.
Despite having low computational complexity in comparison to GFL-NLTV, GFL-ENTV shows improved performance. We can see that the bright scatterers are clearly distinguishable and the spurious pixels have also been eliminated. 
Thus far, we have considered $J=M$ (see (\ref{eq:rec_mod_comp})). Now, we show the performance results of GFL-ENTV with a reduced number of randomly selected frequency samples, i.e., $J<M$. Figures \ref{fig:fltv_pt75}-\ref{fig:fltv_pt25} show the performance of GFL-ENTV with $75\%$ ($J=0.75M$), $50\%$ ($J=0.50M$) and $25\%$ ($J=0.25M$) of frequency samples, respectively. We can see a graceful degradation in performance, in comparison to the case $J=M$. Nonetheless, the gains are still substantial.
\section{Conclusions}
\label{sec:concl}
In this paper, we have proposed graph based SAR imaging for improved spatial resolution and denoising. We have proposed the concept of extended neighbourhood to account for irregularity of SAR spatial grid and nonuniformity of reflectivity field.
Experimental results prove that our proposed method outperforms a number of SAR imaging techniques.
\section*{Acknowledgement}
This work has been approved for submission by TASSC-PATHCAD Sponsor, Chris Holmes, Senior Manager Research, Research Department, Jaguar Land Rover, Coventry, UK.
\bibliographystyle{IEEEtran}

\end{document}